\documentstyle[12pt]{article}
\begin{document}
\begin{center}
{\large\bf
Brane-Black Hole Correspondence and Asymptotics of Quantum Spectrum \\}
\vspace{1cm}
{\sc A.A. Bytsenko\footnote{e-mail: abyts@spin.hop.stu.neva.ru}}\\
{\it State Technical University, St. Petersburg, 195251, Russia}\\
{\sc A.E. Goncalves\footnote{e-mail: goncalve@fisica.uel.br}}\\
{\it Departamento de Fisica, Universidade Estadual de Londrina\\
86051-970, Londrina-Parana, Brasil}\\
{\sc S.D. Odintsov\footnote{e-mail: odintsov@quantum.univalle.edu.co. 
On leave from Tomsk Pedagogical University, 634041 Tomsk, Russia}}\\
{\it Departamento de Fisica, Universidad del Valle\\
A.A. 25360, Cali, Colombia}\\

\vspace{2.5cm}
{\bf Abstract}\\
\end{center}

We discuss  the asymptotic properties of quantum states density for
fundamental (super) membrane in the semiclassical approach. The matching of
BPS part of spectrum for superstring and supermembrane gives the possibility
to get stringy results via membrane calculations and vice versa.
The brane-black hole correspondence (on the level of black hole states and
brane microstates) is also studied.

\vspace*{2cm}

{\bf 1}. It has been realized recently that there are very deep connections 
between
fundamental (super) membrane and (super) string theory. In 
particulary, it has
been shown that the BPS spectrum of states for type IIB string on a circle is
in correspondence with the BPS spectrum of fundamental compactified
supermembrane \cite{schw95-360-13,russ96u-47}.
Remarkable progress has been made towards establishing the string-black
hole correspondence relevant for the extreme black hole \cite{stro96-379-99}. 

The entropy associated with the BPS states is then identical to the 
Bekenstein-Hawking entropy defined by the horizon area. The idea of 
string-black hole correspondence has been formulated also as the 
correspondence 
principle in Ref. \cite{horo96u}.

{\bf 2}. That is the purpose of this work to look to some of these questions 
from fundamental supermembrane point of view. We start from the semiclassical
free energy for fundamental compactified supermembranes (which is known to be
divergent) embedded in flat $D$-dimensional manifolds with topologies
${\cal M}={\bf S}^1\otimes{\bf T}^d\otimes{\bf R}^{D-d-1}$\, $({\bf T}^d$ is
the $d$-dimensional torus). First of all we remind that for simplest
quantum field model the free energy has the form
\cite{byts90-245-21,eliz94}
$$
{\cal F}^{(b,f)}(\beta)=
- \pi^d(\mbox{det}{\cal A})^{1/2}\int_0^{\infty}ds(2s)^{-(D-d+2)/2}\Xi^{(b,f)}
(s,\beta)
$$
$$
\times \Theta\left[\begin{array}{r}
{\bf g}\\
{\bf 0}
\end{array}\right]({\bf 0}|\Omega)
\exp{\left(-\frac{sM^2}{2\pi}\right)}\mbox{,}
\eqno{(1)}
$$
where
$$
\Xi^{(b)}(s,\beta)=\theta_3\left(0|\frac{i\beta^2}{2s}\right)-1\mbox{,}\,\,\,
\,\,\,\,\,\,\,\,\,\, \Xi^{(f)}(s,\beta)=1-\theta_4\left(0|\frac{i\beta^2}{2s}
\right)\mbox{,}
\eqno{(2)}
$$
and $\theta_3(\nu|\tau)$ and $\theta_4(\nu|\tau)=\theta_3(\nu+\frac{1}{2}|\tau)$
are the Jacobi theta functions.
Here ${\cal A}=\mbox{diag}(R_1^{-2},...,R_d^{-2})$ is a $d\times d$ matrix, the
global parameters $R_j$ characterizing the non-trivial topology of ${\cal M}$
appear in the theory owing to the fact that coordinates $x_j (j=1,...,d)$ obey
the conditions $0\leq x_j<2\pi R_j$. The number of topological configurations
of quantum fields is equal to the number of elements in group
$H^1({\cal M};{\bf Z}_2)$, first cohomology group with coefficients in
${\bf Z}_2$. The multiplet ${\bf g}=(g_1,...,g_d)$ defines the topological type
of field (i.e., the corresponding twist), and depending on the field type
chosen in ${\cal M}$, $g_j=0$ or $1/2$. In our case $H^1({\cal M};{\bf Z}_2)
={\bf Z}_2^d$ and so the number of topological configurations of real scalars
(spinors) is $2^d$. We follow the notations and treatment of Ref.
\cite{mumf84} and introduce the theta function with characteristics
${\bf a}, {\bf b}$ for ${\bf a},{\bf b}\in{\bf Z}^d$,
$$
\Theta\left[\begin{array}{r}
{\bf a}\\
{\bf b}
\end{array}\right]({\bf z}|\Omega)=\sum_{{\bf n}\in {\bf Z}^d}\exp{\left[
i\pi({\bf n}+{\bf a})\Omega({\bf n}+{\bf a})+i2\pi({\bf n}+{\bf a})({\bf z}+
{\bf b})\right]}\mbox{,}
\eqno{(3)}
$$
in this connection $\Omega=(is/2\pi^2)\mbox{diag}(R_1^2,...,R_d^2)$. The above
method of the free energy calculation admits the subsequent development for
extended objects. We shall assume that free energy is equivalent to a sum of
the free energies of quantum fields which  present in the modes of a membrane.
The factor $\exp(-sM^2/2\pi)$ in Eq. (1) should be understood as
$\mbox{Tr}\exp(-sM^2/2\pi)$, where $M$ is the mass operator of membrane and the
trace is over infinite set of Bose-Fermi oscillators $N^{(b)}_{{\bf n}},
N^{(f)}_{{\bf n}}$.

{\bf 3}. For the noncompactified supermembrane the question of reliability of
the semiclassical approximation is not absolutely clear \cite{duff88-297-515}.
 The discrete part of the supermembrane spectrum
propagating in eleven-dimensional Minkowski space-time can be written in the
form (see for detail Refs. \cite{duff88-297-515,byts90-245-21})
$$
M^2=\sum_{j=1}^8\sum_{{\bf n}\in{\bf Z}^2/{\{\bf 0}\}}\omega_{{\bf n}}\left(
N_{{\bf n}j}^{(b)}+N_{{\bf n}j}^{(f)}\right)\mbox{,}
\eqno{(4)}
$$
where
$$
\omega_{{\bf n}}=\sqrt{(n_1\pi/a)^2+(n_2\pi/b)^2}\mbox{,}
\eqno{(5)}
$$
and $a=\pi R_1,\, b=\pi R_2$. Thus as a result we have
$$
\mbox{Tr}\sum_{{\bf n}\in {\bf Z}^d/{\{\bf 0}\}}\exp\left(
-\frac{s}{2\pi}M^2\right)
=\left[H_+(\Omega)H_-(\Omega)\right]^8\mbox{,}
\eqno{(6)}
$$
where
$$
H_{\pm}(\Omega)=\prod_{{\bf n}\in{\bf Z}^d/{\{\bf 0}\}}\left\{1\pm\exp[
-({\bf n},\Omega{\bf n})^{1/2}]\right\}^{(\pm 1)}\mbox{,}
\eqno{(7)}
$$
and $\Omega=(s^2/4)\mbox{diag}(a^{-2},b^{-2})$.

For generating functions $H_{\pm}(\Omega),\: \Omega=z\mbox{diag}(1,...,1),
(z=t+2\pi x)$ in the half-plane $\Re z>0$ there exists an asymptotic expansion
uniformly in $x$ as $t\mapsto 0$, provided $|\mbox{arg}z|\leq
\frac{\pi}{4}$ and $|x|\leq\frac{1}{2}$ and given by \cite{byts96u}
$$
H_{+}(\Omega)=\exp\left\{[A\Gamma(p)\zeta_{-}(1+p)z^{-p}-Z_p(0)\mbox{log}2+
O\left(t^{c_{+}}\right)]\right\}\mbox{,}
\eqno{(8)}
$$
$$
H_{-}(\Omega)=\exp\left\{[A\Gamma(p)\zeta_{+}(1+p)z^{-p}-Z_p(0)\mbox{log}z+
Z_p^{'}(0)+O\left(t^{c_{-}}\right)]\right\}\mbox{,}
\eqno{(9)}
$$
where $0<c_{+},c_{-}<1$ and $Z_p(s)\equiv Z_p\left|_{\bf h}^{\bf g}
\right|(s)$ is the $p$-dimensional Epstein zeta function which has a pole with
residue $A$. In above equations $\zeta_{-}(s)\equiv\zeta_R(s)$ is the Riemann
zeta function, $\zeta_{+}(s)=(1-2^{1-s})\zeta_{-}(s)$.
The total number of quantum states can be described by the quantities
$r_{\pm}(N)$ defined by
$$
K_{\pm}(t)=\sum_{N=0}^{\infty}r_{\pm}(N)t^N \equiv H_{\pm}(\Omega)\mbox{,}
\eqno{(10)}
$$
where $t=\exp(-z), t<1$, and $N$ is a total quantum number. By means of the
asymptotic expansion of $K_{\pm}(t)$ for $t\mapsto 1$, which is equivalent to
expansion of $H_{\pm}(\Omega,0)$ for small z and using the formulae (8) and
(9) one arrives at complete asymptotic of $r_{\pm}(N)$. Thus for
$N\mapsto \infty$ one has \cite{eliz94,byts96u}
$$
r_{\pm}(N)=C_{\pm}(p)N^{(2 Z_p(0)-p-2)/(2(1+p))}
$$
$$
\times \exp\left\{\frac{1+p}{p}[A\Gamma(1+p)\zeta_{\pm}(1+p)]^
{1/(1+p)}N^{p/(1+p)}\right\}[1+O(N^{-\kappa_{\pm}})]\mbox{,}
\eqno{(11)}
$$
$$
C_{\pm}(p)=[A\Gamma(1+p)\zeta_{\pm}(1+p)]^{(1-2q Z_p(0))/(2p+2)}
\frac{\exp(Z_p'(0))}{[2\pi(1+p)]^{1/2}}\mbox{,}
\eqno{(12)}
$$
$$
\kappa_{\pm}=\frac{p}{1+p}\min \left(\frac{C_{\pm}(p)}{p}-\frac{\delta}{4},
\frac{1}{2}-\delta\right)\mbox{,}
\eqno{(13)}
$$
and $0<\delta<\frac{2}{3}$.\\

{\bf 4}. Let us consider the membrane excitation states with non-trivial
winding number around the target space torus. In this case the spectrum of the
light-cone membrane Hamiltonian is discrete \cite{duff88-297-515,
byts90-245-21}. The toroidal membrane is wrapped around the target space torus
$({\cal M}={\bf S}^1\otimes{\bf T}^2\otimes{\bf R}^8)$ and the winding number
associated with this wrapping is $l_1l_2$. For $l_1l_2\neq 0$ a membrane is
topologically protected against usual supermembrane instabilities
\cite{wit89-320-135}. In the semiclassical approximation 
\cite{duff88-297-515} the eleven-dimensional mass formula has the form $(d=2)$
$$
M^2(l)=(l_1l_2R_1R_2)^2+ {\cal H}\mbox{,}
\eqno{(14)}
$$
where the oscillator Hamiltonian  can be written as
follows
$$
{\cal H}=2\sum_{{\bf n}\in{\bf Z}^2/{\{\bf 0}\}}
\left(\alpha_{{\bf n}}^{\dag}\alpha_{{\bf n}}+\omega_{{\bf n}}
\sigma_{{\bf n}}^{A\dag}
\sigma_{{\bf n}}^A\right)\mbox{,}
\eqno{(15)}
$$
in addition $\omega_{{\bf n}}^2=\sum_j(l_jn_jR_j)^2$, and $A = 1,...8$ are
$SO(7)$ spinor indices. The two constraints for the toroidal membrane are
$$
C_j=l_jk_j+{\cal N}_{n_j}^{(b)}+{\cal N}_{n_j}^{(f)}=0\mbox{,}
\eqno{(17)}
$$
where $k_j=R_jp_j\, (j=1,2),\, p_j$ are discrete momenta. The commutation
relations for above operators can be found, for example, in Ref.
\cite{duff88-297-515}. The Fock vacuum $|0>$ and the mass of state (obtained
by acting on the vacuum with creation operators) can be written
correspondingly as follows
$$
(\mbox{mass})^2 \sim \alpha_{n_1n_2}^{\dag}...\alpha_{n_in_j}^{\dag}
\sigma_{n_1n_2}^{\dag}...\sigma_{n_in_j}^{\dag}|0>\mbox{.}
\eqno{(19)}
$$
As usual the physical Hilbert space consists of all Fock space states
obeying the conditions (17). The quantized momenta $p_1$ and $p_2$ correspond
to central charges in $N=2$ nine-dimensional supersymmetry that classify the
fluctuations about classical solution. The fact that the $M^2(l)$ is
non-vanishing means that no multiplet-shortening will take place and so the
vacuum must correspond to a long massive $N=2$ multiplet with $M^2(l)=
(l_1l_2R_1R_2)^2$.

Let us demonstrate a correspondence between the semiclassical membrane and
string results. If $R_2\mapsto\infty$ while at the same time increasing $l_2$
such that the product $l_2R_2$ is kept fixed ($l_2R_2=1$, for example) then
one dimension is shrunk while keeping the energy constant and producing
a closed string. In this limit non-zero momentum $p_2$ is excluded and one
obtains from Eq. (14) the mass formula for a ten-dimensional superstring
compactified on a circle of radius $R_1$. Thus the nine-dimensional mass is
given by
$$
M^2=l_1^2R_1^2+\frac{k_1^2}{R_1^2}+2\sum_{j=1}^8\sum_{n_1\neq 0}\left(
\alpha_{nj}^{\dag}\alpha_{nj}+|n_1|\sigma_{nj}^{\dag}\sigma_{nj}\right)
\mbox{.}
\eqno{(20)}
$$
In this limit one of the constraints (17) becomes empty, while the other one
yields
$$
C_1=l_1k_1+\sum_{j=1}^8\sum_{n_1\neq 0}\left(\mbox{sign}(n_1)\alpha_{nj}^{\dag}
\alpha_{nj}+n_1\sigma_{nj}^{\dag}\sigma_{nj}\right)=0\mbox{.}
\eqno{(21)}
$$
Note that in accordance with our definitions $\alpha_n^{\dag}$ and $\sigma_n^
{\dag}$  create left (right) - moving states when $n_1>0\, (n_1<0)$, while
$\alpha_n$ and $\sigma_n$  annihilate left (right) - moving states when
$n_1>0 \, (n_1<0)$. The constraint (21) is equivalent to the usual condition
relating left and right Hamiltonians for the string.

Let us compare the BPS part of the membrane spectrum with the type II string
BPS spectrum. The correspondence between spectra at zero-mode level was
established in Ref. \cite{schw95-360-13}.
Using the constraint (21) in the mass formula (20) for free string states
(note that under T-duality relating IIA and IIB spectra, $R_1\mapsto
\alpha'R_1^{-1}$, for the sake of simplicity here and in the following we
assume an inverse string tension parameter $\alpha'$ equal to 1) we have
$$
M^2=\left(l_1R_1+\frac{k_1}{R_1}\right)^2\mbox{.}
\eqno{(22)}
$$
The Kaluza-Klein mass for perturbative or $(1,0)$ string states with $k_2=0$
(zero Ramond-Ramond charge) is given by
$$
M_{IIB}^2=\frac{k_1^2}{R_1^2}+l_1^2R_1^2+2\left({\cal N}^{(b)}+{\cal N}^{(f)}
\right)\mbox{.}
\eqno{(23)}
$$
For the general BPS perturbative states with the corresponding oscillating
states $(1,0)$ the masses given in formula (23) coincide with the masses of
Eq. (22). The last statement is correct also for the non-perturbative type IIB
string \cite{russ96u-47,russ97u-188} with charges $(q_1,q_2)$. In this case
$k_1=mq_1, k_2=mq_2$, and $q_1,q_2$ being co-prime. For the NS-NS string
we have $k_2=0, q_1=1, q_2=0$, while for the R-R string, $k_1=0, q_1=0,
q_2=1$ and the value of $M_{IIB}^2$ coincides with masses (up to change
of indices $1\mapsto 2$) given in formula (22). Hence, for calculation
of BPS membrane Hagedorn density one can use stringy results.

If $l_2=0$ (or $l_1=0$) then the stable classical solution will be collapsed
to string-like membrane wound around only one compact direction in the target
space. For this classical configuration the mass formula (14) indicates the
presence of massless states. The world-volume metric for this light-cone
classical solution is degenerate, but the field equations are neverthelless
non-singular. It is not clear whether the semiclassical approximation can be
trusted for such a configuration \cite{duff88-297-515}. Indeed the equation
$\omega_{n_1}=\sqrt{(n_1l_1R_1)^2}$ shows that all values of $n$ give the same
frequency $\omega_{n_1}$ in the semiclassical approximation. This infinite
degeneracy will presumably be lifted when the higher order terms in the
Hamiltonian are included.\\

{\bf 5}. Let us suppose that the semiclassical approach to quantization leads
to a second quantized brane theory which can be considered as a theory of
non-interacting strings. Then the Hilbert space of all multiple string states
that satisfy the BPS conditions (zero branes) with a total energy momentum $P$
has the form \cite{dijk96u-26}
$$
{\cal H}_{P}=\bigoplus_{\scriptstyle \sum lN_{l}=N_{P}}\bigotimes_{l}
\mbox{Sym}^{N_{l}}{\cal H}_{l}\mbox{,}
\eqno{(24)}
$$
where symbol $\mbox{Sym}^N$ indicates the $N$-th symmetric tensor product. The
exact dimension of ${\cal H}_{P}$ is determined by the character expansion
formula
$$
\sum_{N_{P}}\mbox{dim}{\cal H}_{P}q^{N_{P}}\simeq\prod_{l}\left(\frac{1+q^l}
{1-q^l}\right)^{\frac{1}{2}\mbox{dim}{\cal H}_l}\mbox{,}
\eqno{(25)}
$$
where the dimension ${\cal H}_l$ of the Hilbert space of single string BPS
states with momentum $k=l\hat{P}$ is given by
$\mbox{dim}{\cal H}_l=d(\frac{1}{2}l^2\hat{P})$, and $|\hat{P}|^2=
|\hat{P}_{L}|^2-|\hat{P}_{R}|^2$ (see Ref. \cite{dijk96u-26} for detail). The
asymptotics of the generating function (25) and the dimension ${\cal H}_{P}$
can be found with the help of Eqs. (11), (12) that is generalization of the
Meinardus result for vector-valued functions.

The Eq. (25) is similar to the denominator formula of a (generalized) Kac-Moody
algebra \cite{borc95-120-161}, which can be written as follows
$$
\sum_{\sigma\in W}\left(\mbox{sgn}(\sigma)\right)e^{\sigma(\rho)}=
e^{\rho}\prod_{r>0}\left(1-e^r\right)^{\mbox{mult}(r)}\mbox{,}
\eqno{(26)}
$$
where $\rho$ is the Weyl vector, the sum on the left hand side is over all
elements of the Weyl group $W$, the product on the right side runs over all
positive roots (one has the usual notation of root spaces, positive roots,
simple roots and Weyl group, associated with Kac-Moody algebra) and each
term is weighted by the root multiplicity $\mbox{mult}(r)$.
The Eq. (25) reduces to the standard superstring partition function for
$\hat{P}^2=0$ \cite{dijk96u-26}. The equivalent description of the second
quantized string states on the five-brane can be obtained, for example, by
considering the sigma model on the target space $\sum_{N}\mbox{Sym}^NT^4$.
There is the correspondence between the formula (24) and the term at order
$q^{\frac{1}{2}N_p{\hat{P}}^2}$ in the expansion of the elliptic genus of the
orbifold $\mbox{Sym}^{N_{P}}T^4$ \cite{dijk96u-26}. Using this correspondence
one finds that the asymptotic growth is equal that of states at level
$\frac{1}{2}N_P{\hat{P}}^2$ in a unitary conformal field theory with central
charge proportional to $N_{P}$.

Our aim now is to compare the quantum states of a membrane and a black hole.
The correspondence between asymptotic state density of the fundamental $p$-brane
and the four-dimensional black hole for $p\to\infty$ was found in Refs.
\cite{byts93-304-235}. An extreme black hole states can be
identified also with the highly excited states of a fundamental string. The
relevant string states are BPS states, while the string entropy is identical
to the Bekenstein-Hawking entropy defined by the horizon area
\cite{stro96-379-99}.
For a calculation of the ground state degeneracy of systems with quantum
numbers of certain BPS extreme black holes the $D$-brane method can be used.
A typical 5-dimensional example has been analyzed in Refs.
\cite{call96-475-645,mald96-475-679,haly96u-12}. Working in the type IIB string
theory on $M^5\bigotimes{\bf T}^5$ one can construct a $D$-brane configuration
such that the corresponding supergravity solutions describe 5-dimensional black
holes. In addition five branes and one branes are wrapped on ${\bf T}^5$ and
the system is given by Kaluza-Klein momentum $N$ in one of the directions.
Therefore the three independent charges $(Q_1,Q_5,N)$ arise in the theory,
where $Q_1$, $Q_5$ are  electric and a magnetic charges respectively. The naive
$D$-brane picture gives the entropy in terms of partition function $H_{\pm}(z)$
for a gas of $Q_1Q_5$ species of massless quanta. For $p=1$ the integers
$r_{\pm}$ in Eq. (11) represent the degeneracy of the state with momentum
$N$. Thus for $N\mapsto\infty$ one has
$$
\mbox{log}r_{\pm}(N)=\sqrt{q\zeta_{\pm}(2)N}-\frac{q+3}{4}\mbox{log}N
+ \mbox{log}C_{\pm}(1) + O(N^{-\kappa_{\pm}})\mbox{,}
\eqno{(27)}
$$
where
$$
C_{+}(1)=2^{-\frac{1}{2}}\left(\frac{q}{16}\right)^{\frac{q+1}{2}}\mbox{,}
\hspace{0.3cm} C_{-}(1)=C_{+}(1)\left(\frac{4}{3}\right)^{\frac{q+1}{2}}\mbox{.}
\eqno{(28)}
$$
For fixed $q=4Q_1Q_5$ the entropy is given by
$$
S=\left[\mbox{log}\left(r_{+}(N)r_{-}(N)\right)\right]_{N\mapsto\infty}
\simeq c_1\sqrt{Q_1Q_5N}-\mbox{log}N(c_2Q_1Q_5+c_3)\mbox{,}
\eqno{(29)}
$$
where $c_j\, (j=1,2,3)$ are some positive numbers. This expression agrees with
the classical black hole entropy.

Note that for $p>1$, using the mass formula $M^2=N$, for the
number of branes states of mass $M$ to $M+dM$ one can find (see also
\cite{byts93-304-235})
$$
\varrho(M)dM\simeq 2C_{\pm}(p)M^{\frac{p-1}{p+1}}\exp\left[b_{\pm}(p)
M^{\frac{2p}{p+1}}\right]\mbox{,}
\eqno{(30)}
$$
$$
b_{\pm}(p)=(1+\frac{1}{p})\left[A\Gamma(p+1)\zeta_{\pm}(p+1)
\right]^{\frac{1}{p+1}}\mbox{.}
\eqno{(31)}
$$
This result has a universal character for all p-branes.

Recently it has been pointed out that the classical result (29) is incorrect
when the black hole becomes massive enough for its Schwarzschild radius to
exceed any microscopic scale such as the compactification radii
\cite{mald96-475-679,haly96u-12}. Indeed, if the charges $(Q_1,Q_5,N)$ tend to
infinity in fixed proportion $Q_1Q_5=Q(N)$, then the correct formula does not
agree with the black hole entropy (29). If, for example, $Q(N)=N$, then for
$N\mapsto\infty$ one finds $\mbox{log}\left[r_{+}(N)r_{-}(N)\right]
\sim N\mbox{log}N$. The naive D-brane prescription, therefore, fails to agree
with U-duality which requires symmetry among charges $(Q_1,Q_5,N)$
\cite{haly96u-12}.


\begin{thebibliography}{9}

\bibitem{schw95-360-13}
J.H. Schwarz, Phys. Lett. {\bf B 360}, 13 (1995).

\bibitem{russ96u-47}
J.G. Russo and A.A. Tseytlin, hep-th/9611047 (1996).

\bibitem{stro96-379-99}
A. Strominger and C. Vafa, Phys. Lett. {\bf B 379}, 99 (1996);
G.T. Horowitz and A. Strominger, Phys. Rev. Lett. {\bf 77}, 2368 (1996);
C.V. Johnson, R.R. Khuri and R.C. Myers, Phys. Lett. {\bf B 378}, 78 (1996).

\bibitem{horo96u}
G.T. Horowitz and J. Polchinski, hep-th/9612146 (1996).

\bibitem{byts90-245-21}
A.A. Bytsenko and S.D. Odintsov, Phys. Lett. {\bf B 245}, 21 (1990);
A.A. Bytsenko and S.D. Odintsov, Fortschr. Phys. {\bf 41}, 233 (1993).

\bibitem{eliz94}
E. Elizalde, S.D. Odintsov, A. Romeo, A.A. Bytsenko and S. Zerbini,
{\em "Zeta Regularization Techniques with Applications"}, World Sci., 
Singapore (1994); A.A. Bytsenko, G. Cognola, L. Vanzo and S. Zerbini, Phys. 
Reports {\bf 266}, No 1,2 (1996).

\bibitem{mumf84}
D. Mumford, {\em "Tata Lectures on Theta I, II"}, Birkh{\"a}user (1983, 1984).

\bibitem{duff88-297-515}
M.J. Duff, T. Inami, C.N. Pope, E. Sezgin and K.S. Stelle, Nucl. Phys. 
{\bf B 297}, 515 (1988).

\bibitem{byts96u}
A.A. Bytsenko and S.D. Odintov, hep-th/9611151 (1996).

\bibitem{wit89-320-135}
B. de Wit, M. L{\"u}scher and H. Nicolai, Nucl. Phys. {\bf B 320}, 135 (1989).

\bibitem{russ97u-188}
J.G. Russo, hep-th/9701188 (1997).

\bibitem{dijk96u-26}
R. Dijkgraaf, E. Verlinde and H. Verlinde, hep-th/9603126 (1996).

\bibitem{borc95-120-161}
R.E. Borcherds, Invent. Math. {\bf 120}, 161 (1995);
J.A. Harvey and G. Moore, Nucl. Phys. {\bf B 463}, 315 (1996).

\bibitem{byts93-304-235}
A.A. Bytsenko, K. Kirsten and S. Zerbini, Phys. Lett. {\bf B 304}, 235 (1993);
A.A. Bytsenko, K. Kirsten and S. Zerbini, Mod. Phys. Lett. {\bf A 9}, 1569 
(1994).

\bibitem{call96-475-645}
C. Callan and J. Maldacena, Nucl. Phys. {\bf B 475}, 645 (1996).

\bibitem{mald96-475-679}
J. Maldacena and L. Susskind, Nucl. Phys. {\bf B 475}, 679 (1996).

\bibitem{haly96u-12}
E. Halyo, A. Rajaraman and L. Susskind, hep-th/9605112 (1996).

\end{thebibliography}
\end{document}